\documentclass[a4paper,11pt]{article}
\usepackage{pos}

\newcommand{\Eq}[1]{Eq.~(\ref{#1})}

\newcommand{\Fig}[1]{Fig.~\ref{#1}}

\title{Quantum thermalization of gauge theories:\\chaos, turbulence and universality}
\author*[a,b]{Niklas Mueller}
\author[c,d]{Torsten V. Zache}
\author[e]{Robert Ott}

\affiliation[a]{Maryland Center for Fundamental Physics and Department of Physics, University of Maryland,\\College Park, MD 20742, USA.}
\affiliation[b]{Joint Quantum Institute, NIST/University of Maryland, College Park, MD 20742, USA. }
\affiliation[c]{Center for Quantum Physics, University of Innsbruck, 6020 Innsbruck, Austria}
\affiliation[d]{Institute for Quantum Optics and Quantum Information of the Austrian Academy of Sciences,\\6020 Innsbruck, Austria}
\affiliation[e]{Heidelberg University, Institut f\"{u}r Theoretische Physik, Philosophenweg 16, 69120 Heidelberg, Germany}

\emailAdd{niklasmu@umd.edu}

\abstract{In this talk, we discuss real-time thermalization dynamics of $\mathbf{Z}_2$ Lattice Gauge Theory in 2+1 spacetime dimensions~\cite{mueller2021thermalization}. While classical thermalization is commonly associated with chaotic behavior, turbulence
and universality, the manifestation of these phenomena in quantum mechanical systems is not clear.  However, when viewed through the lens of Entanglement Structure, we find that quantum thermalization proceeds in characteristic stages and reveals  phenomena remarkably similar to their classical counterparts: chaos, turbulence and universality.
}

\FullConference{%
 The 38th International Symposium on Lattice Field Theory, LATTICE2021
  26th-30th July, 2021
  Zoom/Gather@Massachusetts Institute of Technology
}

\begin{document}
\maketitle
%
%
%	INTRODUCTION
%
\section{Introduction}
Recent advances in simulating quantum many-body systems with digital quantum computers and analog devices, based on atomic, molecular and optical (AMO) systems, have opened new avenues to address old problems~\cite{cirac2012goals,hauke2012can,martinez2016real,klco2018quantum,zache2018quantum,davoudi2020towards,ciavarella2021trailhead,Wiese2021}. One such question  
is the thermalization of gauge theories, relevant e.g. for Quantum Chromodynamics (QCD) in ultra-relativistic heavy ion collisions~\cite{berges2020thermalization}, and in many other fields ranging from atomic gases~\cite{schachenmayer2015thermalization}, to condensed matter physics~\cite{nandkishore2015many}, and cosmology~\cite{micha2004turbulent}. 

Much understanding has been derived from the Eigenstate Thermalization Hypothesis~\cite{deutsch1991quantum,srednicki1994chaos} and it has become clear that entanglement is an important ingredient in thermalization, yet the latter is barely explored for gauge theories because of its ambiguous definition~\cite{buividovich2008entanglement,casini2014remarks,aoki2015definition,radivcevic2016entanglement}. In this work, we overcome this issue for $\mathbf{Z}_2$ LGT in (2+1) spacetime dimensions ($\mathbf{Z}_2^{2+1}$), by developing dual formulations  `with entanglement cuts', allowing us to compute the Entanglement Structure of non-equilibrium states. Focusing on quench dynamics of an initial unentangled state, we investigate the `Entanglement Spectrum' (ES), a representation of a state in terms of an `Entanglement Hamiltonian' (EH), analogous to energy levels, first suggested by Li and Haldane as an indicator of topological order in non-Abelian fractional Quantum Hall effect systems~\cite{li2008entanglement}.

\begin{figure}[t]
\begin{center}
\includegraphics[width=0.75\textwidth]{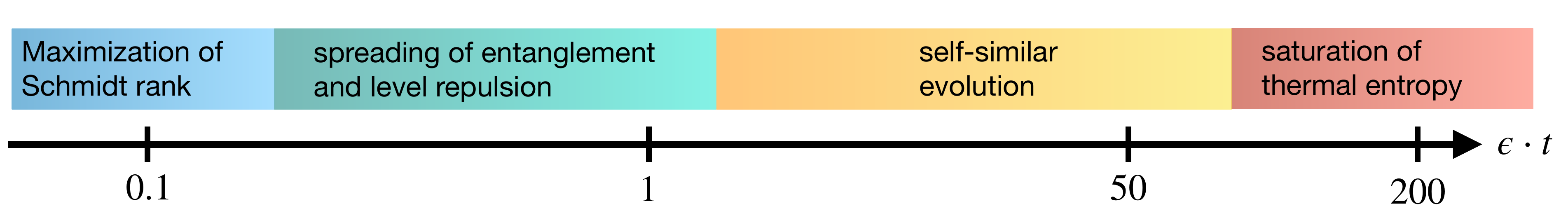}
\caption{Overview of the stages of quantum thermalization of $\mathbf{Z}_2^{2+1}$, including (exponential) growth of Schmidt values and build-up of level repulsion at earliest time, and saturation of the von-Neumann entropy at a parametrically later stage. An intermediate regime is characterized by self-similar evolution, typical for (classical) wave turbulence.}
\label{fig:overview}
\end{center}
\end{figure}

We find that thermalization proceeds in clearly separated stages, c.f. \Fig{fig:overview}: Exponentially-fast growth of Schmidt values and maximization of the rank of the reduced density matrix at earliest times, followed by spreading of ES level repulsion, and saturation of entanglement entropy at parametrically later times. An intermediate regime is characterized by self-similar evolution of the Schmidt spectrum, with scaling coefficients $\alpha=0.8\pm 0.1$ and $\beta = 0.05 \pm 0.03$, reminiscent of classical wave turbulence and universality.   

%
%
%	HAMILTONIAN FORMULATION
%
\section{Hamiltonian Formulation of $\mathbf{Z}_2^{2+1}$ Lattice Gauge Theory}
The Hamiltonian of $\mathbf{Z}_2^{2+1}$ LGT is~\cite{mueller2021thermalization,wegner1971duality}
\begin{align}\label{eq:Z2Hamiltonian}
H= - \sum_{\mathbf{n}} \sigma^z_{\mathbf{n},x}\sigma^z_{\mathbf{n}+\hat{x},y}\sigma^{z}_{\mathbf{n}+\hat{y},x}\sigma^z_{\mathbf{n},y} -\epsilon\sum_{\mathbf{n},i=x,y} \sigma^x_{\mathbf{n},i} \,,
\end{align}
where $\sigma^{x/z}_{\mathbf{n},i}$ are Pauli operators positioned along the links of a two-dimensional spatial lattice $\mathbf{n}\equiv (n_x,n_y)$ with $n_i \in [0,N_i-1]$. Gauss law $G_\mathbf{n} \equiv  \prod_i \sigma^x_{\mathbf{n}}\sigma^x_{\mathbf{n}-\hat{i}}$  specifies the physical sector $G_\mathbf{n} | \psi^{\rm phys} \rangle = | \psi^{\rm phys}\rangle $. $\mathbf{Z}_2^{2+1}$ LGT was first proposed by Wegner \cite{wegner1971duality} as a model containing a phase transition without a local order parameter, a deconfinement ($\epsilon < \epsilon_c$) versus confinement ($\epsilon > \epsilon_c$) transition, later understood in the context of topological order (TO) \cite{sachdev2018topological}. 
Our main interest is the Entanglement Spectrum $\{\xi_\lambda\}$ of a quantum state $|\psi\rangle$, defined from its Schmidt decomposition
\begin{align}
|\psi \rangle = \sum{}_{\lambda=1}^\chi \, e^{- \frac{\xi_\lambda}{2}} | \phi^\lambda_A \rangle | \phi^\lambda_B \rangle\,,
\end{align}
where $| \phi^\lambda_A \rangle$ and  $| \phi^\lambda_B \rangle$ are the Schmidt vectors of a bipartition into subsystems $A$ and $B$, respectively, and $\chi$ the rank of the reduced density matrices,
\begin{align}
\rho_A = \sum{}_{\lambda=1}^\chi   \, e^{- {\xi_\lambda}} | \phi^\lambda_A \rangle \langle  \phi^\lambda_A | \equiv \exp(-H^{\rm ent.})\,,
\end{align}
which defines the Entanglement Hamiltonian (EH), $H^{\rm ent.}$  ($\chi=1$ denotes an un-entangled state).

A duality, between \Eq{eq:Z2Hamiltonian} and a two-dimensional Ising model, for an infinite system was recognized in \cite{wegner1971duality,horn1979hamiltonian}, however, the naive definition of entanglement on the Ising-side of the duality does not match that of the gauge theory.  In light of this, we developed generalized dualities~\cite{mueller2021thermalization}, 
whereby the dual theory is embedded into a larger, unphysical Hilbertspace along entanglement cuts and physical boundaries,
enabling us to  compute
the entanglement spectrum of the gauge theory. 

We  verify the validity of our approach by explicitly demonstrating Li and Haldane's bulk-boundary conjecture~\cite{li2008entanglement}, for the first time in a gauge theory, both analytically in perturbation theory for an infinite systems, as well as numerically on a torus at arbitrary coupling~\cite{mueller2021thermalization}. We also reconstruct the Entanglement Hamiltonian, using an ansatz based on the Bisognano-Wichmann theorem~\cite{bisognano1975duality,bisognano1976duality}, and find good agreement with expectations from the theorem~\cite{zache2021entanglement} in the topological ordered and trivial phase, as well as at the critical point $\epsilon=\epsilon_c$.
Using the closure of the `Entanglement Gap' \cite{li2008entanglement} of the ES as an indicator, we can determine the critical coupling $\epsilon_c$ to within 10\% accuracy~\cite{blote2002cluster}, even on very small lattices $N_x \times N_y = 6\times 4$~\cite{mueller2021thermalization}.
%
%
%	THERMALIZATION
%
\section{Thermalization of $\mathbf{Z}_2^{2+1}$ Lattice Gauge Theory}
Considering the entanglement of a bipartition of a two-dimensional lattice with periodic boundary conditions,  we study non-equilibrium dynamics after a quench by computing the ES of non-equilibrium states. Towards this end, we prepare an initially unentangled state, a random (excited) eigenstate of \Eq{eq:Z2Hamiltonian} for $\epsilon \rightarrow \infty$, and then time-evolve it with the full Hamiltonian, for $\epsilon =1$.  

\begin{figure}[t]
\begin{center}
\includegraphics[width=0.95\textwidth]{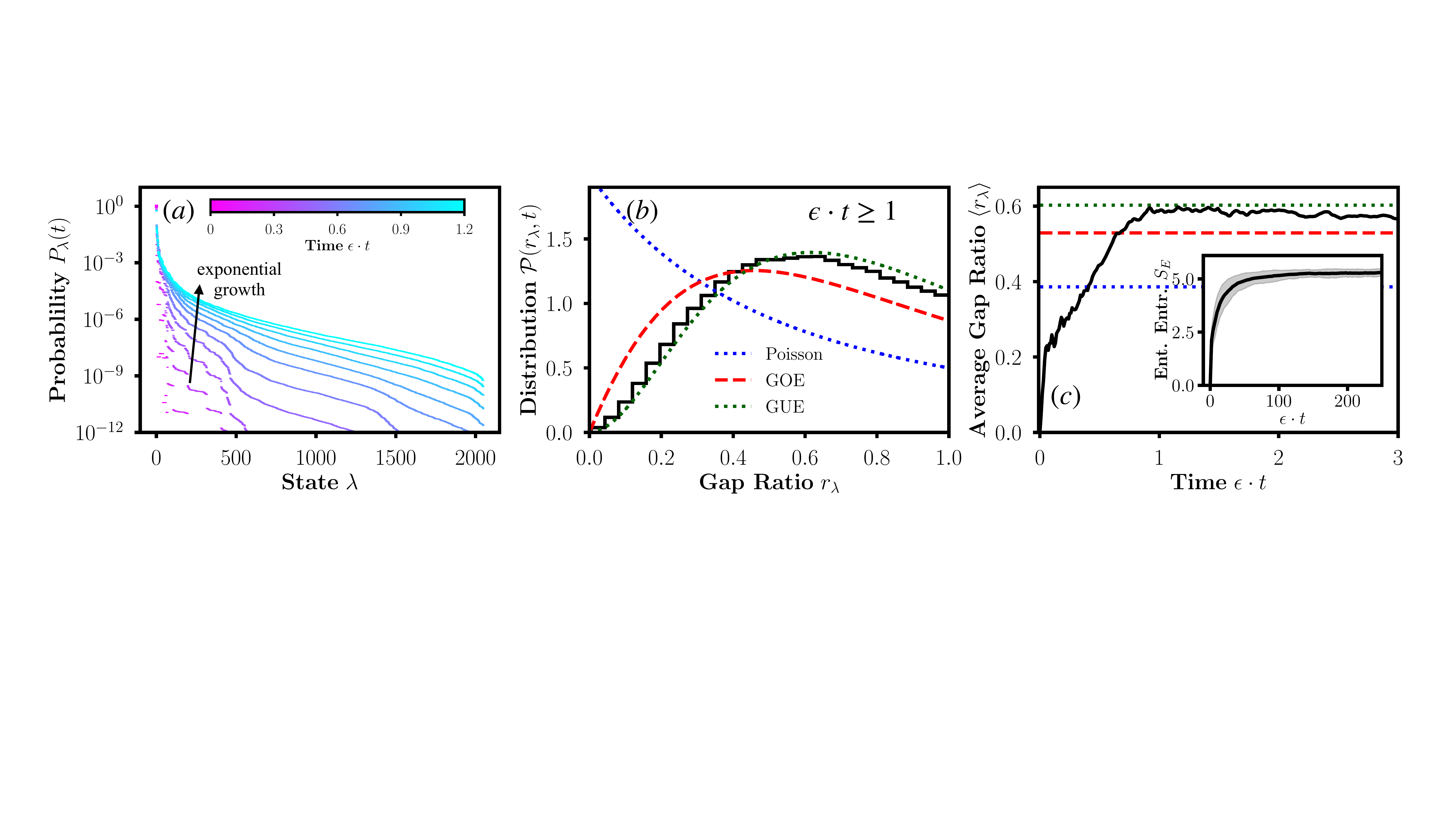}
\caption{(a) Exponential growth in the Schmidt spectrum $P_\lambda(t)$, in the early-time regime $\epsilon \cdot t \le 1$.  Level repulsion: (b) distribution of gap ratio $r_\lambda$ for $\epsilon \cdot t \ge 1$ and (c) the level distribution is consistent with GUE and chaotic behavior at late times. Inset of (c): Saturation of the von Neumann entropy $S_E$ to thermal entropy at asymptotically late times $\epsilon \cdot t \ge 150$.  Results are for a lattice of $N_x\times N_y = (3+5)\times 3$. }
\label{fig:plot1}
\end{center}
\end{figure}

We find that the thermalization process occurs in characteristic stages with clearly separated timescales, depicted in \Fig{fig:overview}.
At earliest times, $\epsilon \cdot t \lessapprox 1$, there is quick change from an initially un-entangled product state, with Schmidt rank   $\chi=1$, to a state with maximal rank $\chi= \text{dim} (A)$.  As is illustrated in \Fig{fig:plot1}(a), this phase is characterized by exponential growth of Schmidt values ${P}_\lambda(t)\equiv \exp{(- \xi_\lambda(t) )}$,  reminiscent of unstable mode growth  in classical plasmas \cite{nazarenko2011wave}.

We study the evolution of the system from the perspective of level statistics of the ES, an important measure indicating the presence of chaos in the system~\cite{borgonovi2016quantum}, by computing the level spacing distribution of the unfolded ES ~\cite{guhr1998random} and the gap ratio~\cite{oganesyan2007localization},
\begin{align}
r_\lambda\equiv \frac{\min(\delta_\lambda,\delta_{\lambda-1})}{\max(\delta_\lambda,\delta_{\lambda-1})}\,,
\end{align}
where $\delta_\lambda\equiv \xi_\lambda-\xi_{\lambda-1}>0$ are the level spacings of the ES. \Fig{fig:plot1}(b) shows the distribution of gap ratios, $\mathcal{P}(r_\lambda,t)$ for $\epsilon\cdot t\ge 1$ and \Fig{fig:plot1}(c) its time dependence, with average $\langle r \rangle^{\rm Poisson} \approx 0.38$, $\langle r \rangle^{\rm GOE} \approx 0.52$ and $\langle r \rangle^{\rm GUE} \approx 0.60$. We compare to expectations 
from a completely uncorrelated Poisson distribution (blue dotted), commonly associated with many-body localization, versus Gaussian Orthogonal (GOE, red dashed) and Gaussian Unitary Ensembles (GUE, green dotted) attributed to the presence of chaotic behavior\footnote{Because contributions from different symmetry sectors are always uncorrelated, we consider  the symmetry-resolved ES by projecting the reduced density matrix into its symmetry sectors.}.
The EH is consistent with  GUE statistics at $\epsilon  \cdot t\ge 1 $ and remains so at all times, we note that the level spacing statistics of the physical Hamiltonian \Eq{eq:Z2Hamiltonian} for $\epsilon =1$ is GOE\footnote{The significance of GUE over GOE is the absence of time reversal invariance. While one 
may naively assume that the EH approaches \Eq{eq:Z2Hamiltonian}, in line with a Gibbs ensemble i.e. $\rho_A^{\rm thermal} \sim \exp{(-\beta H_A)}$, and that the level statistics should be equal, we point out that this is not exactly correct because it should be the global state corresponding to a Gibbs ensemble, i.e. $\rho_A^{\rm thermal} = \text{tr}_B (\rho^{\rm can.}) $, where $\rho^{\rm can.} = \exp(-\beta H)/\text{tr} (\exp(-\beta H))$~\cite{garrison2018does}. }.

\begin{figure}[t]
\begin{center}
\includegraphics[width=0.5\textwidth]{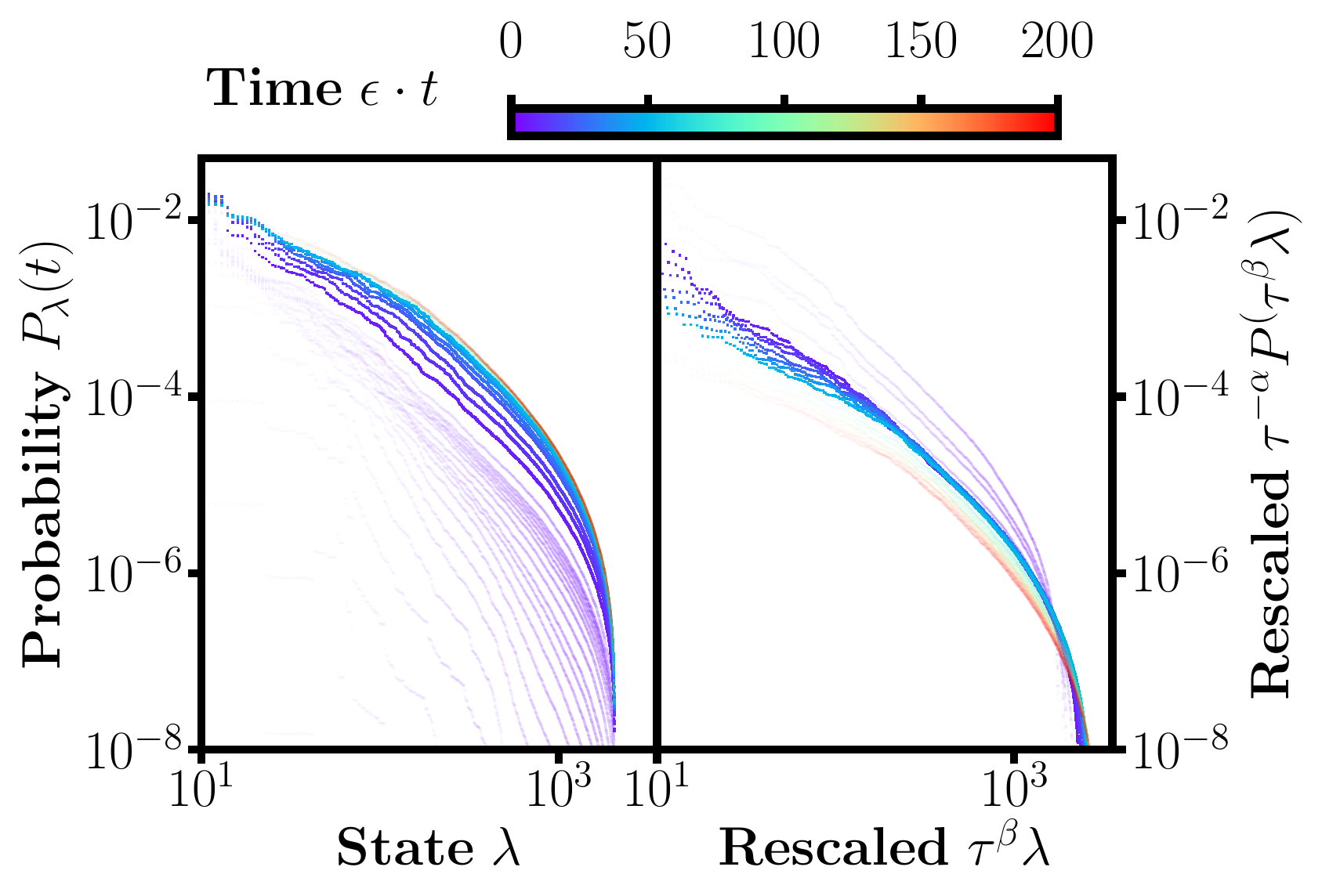}
\caption{Left: Un-rescaled Schmidt spectrum $P_\lambda(t)$ for different times. Right: The rescaled spectrum is characterized by a self-similar universal form in the scaling regime, spanning roughly $\epsilon \cdot  t \approx 2-80$. The spectra for times outside this scaling regime, $\epsilon \cdot t \le 2$ and $\epsilon \cdot t \ge 80$, are shaded out.}
\label{fig:plot2}
\end{center}
\end{figure}

At intermediate times $\epsilon \cdot t \approx 2 -80$, the approach to thermal equilibrium is characterized by a self-similar form of the Schmidt spectrum, see \Fig{fig:plot2},
\begin{align}
P_\lambda(t) = \tau^{-\alpha} P(\tau^\beta \lambda )\,,
\end{align}
where $\tau \equiv \epsilon\cdot(t-t_0)$ with scaling exponents  $\alpha = 0.8\pm 0.1$, $ \beta =0.05\pm 0.03 $
and $\epsilon\cdot t_0 = 1.8 \pm 0.2$,  the error reflecting the systematic uncertainty of our statistical analysis but not finite volume effects, which we believe are significant. Self-similar scaling  behavior is characteristic of wave turbulence and universality in classical systems~\cite{nazarenko2011wave},  consequently we call this phenomenon `quantum turbulence'.

We point out that the scaling regime does not cover the whole range $\lambda$, as can be seen in \Fig{fig:plot2}. Whether this is because of the presence of a separate scaling regime in the `low energy part' of the ES, as is typical, e.g. for classical non-Abelian plasmas~\cite{berges2014turbulent,berges2014basin}, or simply due to finite-volume effects is not clear and deserves further investigation. Given the exponential cost for classical computers to do so, this is a strong motivation for exploration with near-future digital quantum computers and analog simulators.
 
Displayed in the inset of \Fig{fig:plot1}(c), the system finally reaches thermal equilibrium at late times $\epsilon \cdot t \ge 150$, marked by the von-Neumann entropy $S_E = - \text{tr}_A (\rho_A \log(\rho_A))$ saturating to the thermal entropy of a global Gibbs state. %, with temperature corresponding to the average energy density of the initial state.

\section{Discussion}
Understanding thermalization of gauge theories is a major motivation for high energy and nuclear physics, e.g., in the context of Quantum Chromodynamics (QCD) probed in ultra-relativistic heavy ion collisions, but also for deeply inelastic scattering experiments at a future Electron-Ion Collider. While thermalization of QCD has been studied intensely, mostly by semi-classical approaches ~\cite{berges2020thermalization}, the implications of Eigenstate Thermalization Hypothesis or Entanglement Structure were not addressed. 
For example, gauge theories have an extensive number of constraints, and one may expect their thermalization dynamics to be different  from non-gauge systems. Our results suggest that this is not the case, and we expect the uncovered 
thermalization scenario to apply generically to many other systems~\footnote{In fact, the large separation in time scales, between build-up of level repulsion and thermalization, has been noted first in non-gauge systems~\cite{rakovszky2019signatures}.}.

Understanding thermalization is an important motivation for exploration of high energy and nuclear physics with quantum computers and analog simulators, using tools such as Entanglement Structure.
Extracting Entanglement Hamiltonians and Entanglement Spectra in state-of-the-art quantum simulator experiments has already been demonstrated using efficient Entanglement Hamiltonian Tomography (EHT) protocols~\cite{pichler2016measurement,dalmonte2018quantum,elben2020mixed,kokail2020entanglement,kokail2021quantum,zache2021entanglement}.  

As shown in \cite{mueller2021thermalization}, Entanglement Structure is also key to understand topological phases, e.g. of condensed matter systems~\cite{kasper2020jaynes,wen2004quantum} and in the context of topological quantum computation~\cite{kitaev2003fault,satzinger2021realizing}, and is
 central to an interdisciplinary interchange of high energy and nuclear physics with these fields.

Clearly, a long and winding road  lies ahead, from $\mathbf{Z}^{2+1}_2$ to quantum simulating QCD. Nonetheless, Entanglement Structure and Eigenstate Thermalization,  may already give important phenomenological insights, e.g., into the rapid hydrodynamization and thermalization of the quark-gluon plasma in ultra-relativistic heavy ion collisions, including (quantum) chaos, turbulence and universality. We will explore those opportunities in the near future. 

\section*{Acknowledgments} N.M. thanks Zohreh Davoudi, Mohammad Hafezi and Christopher White for discussions. N.M. acknowledges funding by the U.S. Department of Energy’s Office of Science, Office of Advanced Scientific Computing Research, Accelerated Research in Quantum Computing program award DE-SC0020312. N.M. also acknowledges support by the U.S. Department of Energy, Office of Science, Office of Nuclear Physics, under contract No. DE-SC0012704
and by the Deutsche Forschungsgemeinschaft (DFG, German Research Foundation) - Project 404640738 during early stages of this work.
R.O. acknowledges funding from the DFG (German Research Foundation) – Project-ID 273811115 – SFB 1225 ISOQUANT.
T.V.Z. thanks Christian Kokail and Peter Zoller for valuable discussions.
T.V.Z.'s work is supported by the Simons Collaboration on Ultra-Quantum Matter, which is a grant from the Simons Foundation (651440, P.Z.).

%---------------------------------------------------------------------------------------------------------------------------
% 		REFERENCES
%---------------------------------------------------------------------------------------------------------------------------    
\bibliographystyle{apsrev4-2} 
\bibliography{references}

\end{document}